# *Online "Advanced Labs" in Physics*

Peter A. Bennett

Physics Department, Arizona State University, Tempe AZ 85287-1504

At Arizona State University we have built the first and only fully online Bachelor of Science degree in Physics, with a complete curriculum, including labs. The upper division Advanced Lab courses present a special challenge for online delivery. We address that using a set of custom-built simulator modules that replicate all the "imperfections" (noise, background, etc) inherent in real-world data. The set of experiments duplicates those of the in-person classes. In this paper, we present an overview of these labs and discuss the advantages and challenges of delivering them online. We assert that these labs provide a valid and rigorous component for the fully online degree. The entire set of labs is available as Open Source Supplemental Materials and is shared for others to use in part or in whole, with suitable attribution (license CC-BY-NC).

## *Introduction*

Online instruction is booming around the world, at all levels from K-12 to post-bachelor and across disciplines. At the university level, most offerings are found in lower-division courses, with relatively few at higher level. That said, many institutions do offer undergraduate degrees in many subjects in online format. Physics is notably different from other sciences and engineering in this regard.[1-4] Indeed, at present there is only one fully online Bachelor of Science degree in Physics in the country, which is offered here at ASU. Our program includes a complete curriculum of fully online courses that matches the in-person program in every aspect, including Advanced Labs. A few programs elsewhere offer partial curricula but not the full degree.[5] The ASU program has enjoyed remarkable success since its launch in Fall 2021, with enrollment already three times that of our in-person program.

The development of fully online Advanced Lab courses presents a special challenge. Indeed, online Advanced Lab might be considered an oxymoron and an insurmountable obstacle to building a full degree program. At ASU we have faced that challenge directly and with good success. We are not aware of any other fully online Advanced Lab courses, at present.

Our online Advanced Lab features custom-built simulator modules that are described below in detail. Simulators and virtual labs have been widely used for online instruction, mostly at the lower-division undergraduate level.[6, 7] A few are available for upper-division, graduate or professional level.[1, 8-10] Commercial products, such as Labster and PhET are available and continue to evolve, but these are exclusively used in introductory level courses.[11] [12, 13] The pedagogic impact of virtual labs has been measured in many settings, and is generally found to improve conceptual understanding and motivation.[6, 7, 14, 15] In an extensive review of 50 empirical studies, it was found that the learning outcome for virtual laboratories were equal or better than that of traditional labs across all categories of learning.[16] In a similar vein, a measurement of physical vs. virtual manipulative experimentation showed no gain for physicality, contradicting a common assumption that hands-on is better and necessary.[14] In these studies, it is noted that the overall assessment depends on the choice and weighting of learning outcomes, which can vary widely. Many studies conclude broadly that virtual labs provide a useful supplement but not a full substitute for in-person labs. An alternative approach for online labs is remote control of physical experiments. A brilliant example is Fygenson, et al. at UCSB.[17]

We note in passing that digital (and remote) user interfaces are increasingly common for modern equipment, such as a digital oscilloscope or lock-in amplifier and for all manner of instrument controls at national facilities (synchrotron light source, free electron laser, neutron scattering, space telescope, etc). We also note that simulators are powerful learning tools, widely used by surgeons, pilots, police, young drivers, etc., providing an important enhancement and acceleration for final on-ground training.

From the comments and references above, it is apparent that the issue of virtual vs. in-person labs is a complex, on-going debate. This paper is not designed to augment that debate with pedagogic data. Rather, we simply provide a "proof of concept" for a rigorous fully online Advanced Lab course. This appears to be the first example in the literature. It follows that there are no pedagogic assessments of such a course, yet. We hope that such studies will follow as others adopt our course in part or in whole.

Advanced Lab courses are found in almost every undergraduate physics program. A comprehensive review of the structure of such courses has recently been published and includes a large bibliography and links to many further resources.[18] A comprehensive listing of desired learning outcomes for Advanced Lab courses is available in an AAPT report.[19, 20] Our online lab classes are carefully built to achieve the learning outcomes of our in-person classes, including:

1. Historical context for Nobel-Prize-winning experiments
2. Choice of parameters for data collection (scan length, step and dwell), resolution, etc.
3. Procedures for assembly and alignment of apparatus
4. Fitting data to model functions, with defined uncertainties
5. Exploration of systematic uncertainties as "what if" scenarios
6. Maintenance of a research notebook
7. Preparation of formal reports with proper organization, content and format

The experiments included in our two in-person Advanced Lab courses are: Young's double-slit diffraction with photon counting, Compton scattering, NMR, Zeeman, Hall effect, interferometry, HeNe laser, X-Ray Physics and quantum entanglement. A video overview tour of these experiments is available at [21]. For in-person classes, students work in groups of 2-3 with each module running 3 weeks. A few extra weeks accommodate topics of uncertainties, data-fitting, electronics recap and writing a formal report. For the online classes students work alone (although collaboration on analysis is encouraged) with the same set of experiments (except Compton and NMR).

The online format for Advanced Lab offers both advantages and disadvantages. Advantages include:

1. Accessibility: Most online students are "adult learners" with heavy work and/or family responsibilities. The flexibility of time and location is critical for them. We believe the high enrollment is driven primarily by accessibility.
2. Diversity: A companion positive attribute is the diversity in the online cohort. For example, we had 34% with military affiliation (past or present) and 30% female in fall 2024.
3. High-end equipment: The simulations replicate the performance of research-grade equipment that is expensive, fragile and dangerous, thus not available at all institutions.
4. Accountability: Every student does every step and is measured with a detailed rubric that provides rapid, structured feedback and builds good technique for data taking, analysis, notebooking and reporting. This does not always occur for in-person programs.

5. Works every time: The simulator produces data with proper setup, alignment, etc which "guarantees" valid/good data in a timely manner. This does not always occur for in-person programs.
6. Lower cost: (see #3 above)

Disadvantages include:

1. No physical interaction with the equipment.
2. No direct group interactions.
3. No design-refine cycle for building and running an experiment.
4. No modeling for interpretation of results (most common in lower-division courses).
5. "Works every time" (see #5 above) could be viewed as a disadvantage, since there is no opportunity to recognize problems and de-bug.

### *Assessement:*

We are not attempting here to provide a measured or comprehensive assessment of the pros and cons of online vs in-person format. Rather, we provide a "proof of concept" for the online format and point to a remarkable success in our own programs at ASU. We have provided Open Source materials for the entire course and hope that instructors elsewhere will implement the course in part or in whole, as appropriate for their own institution. We also hope that some instructors would undertake suitable "pedagogic assessments" to compare online vs in-person format.

We do not claim here that online courses are a superior or direct replacement for in-person Advanced Lab courses. Rather, we assert that the online versions provide valid and rigorous components for an online degree. This is a nuanced point. At ASU, students in the online vs in-person college are completely separated in all programs across the institution, even though the courses themselves are equivalent in content.[22] The striking point is that online enrollment has quickly surpassed in-person enrollment for virtually every degree offered. We believe this demand is driven by excellent quality and accessibility for a diverse student cohort.

We note that the online experiments might usefully be deployed in a hybrid format that includes both online and physical content. This is already happening in our own in-person labs. Alternatively, one might interleave online and physical experiments during the semester, with the aim of including positive features of each mode. Another possibility is that students from another institution might take our entire online course as a non-degree seeking student, if transfer credit is allowed at their home institution.

One method to address items 1 and 2 in the list of disadvantages would be to host an optional campus workshop during the semester break. Students would visit the lab to assemble, align and operate all the equipment, which could be done quickly without the need for "real data", analysis and formal reports. Students would have the opportunity to meet their online cohort, as well as in-person students from the department, as well as the instructor and other faculty. It would be an excellent experience for the students and an excellent recruiting tool for our graduate programs. The campus visit program will be implemented in the near future.

### *Assignments:*

The course is administered via Canvas in asynchronous mode to accommodate multiple time-zones that literally span the globe. We teach it at double-speed (7.5 week session) to improve modularity in the students' scheduling. The pedagogic approach is highly prescriptive, with daily assignments that link content both forward and backward. A few examples are given in the Appendix so the reader may gauge complexity and context. Timely grading (generally within 24 hours) provides critical feedback for future assignments. For example, assignments for background reading (theory, historical and contemporary context), data collection and analysis (fitting to theory including uncertainties) are graded and returned so students can amend their analysis before the formal report. This separation of analysis from reporting allows for detailed grading and feedback on format items of the report (abstract, figures, numbered equations, references, etc). This approach contrasts with typical in-person labs, which often are graded on lab reports alone and may include poor or unusable data and may be done in group style. Ironically, we believe the online format provides better accountability in this respect. Indeed, suitable online components have been adopted in our in-person courses, with good effect. Another component of the online course is the Electronic Lab Notebook (ELN). Students keep remarkably detailed records of their activities, using narrative, images, links, etc. and an indexed Table of Contents. This comprises an excellent item in a student portfolio when they apply for work or graduate studies. They also are required to keep an "Activity Log" showing hours spent on various tasks. This provides useful feedback to instructors for balancing assignment length and difficulty.

***Simulator overview:***

Students obtain data using custom-built simulator modules that produce signals with the full scope of "imperfections" inherent in real measurements including noise (Poisson, 1/f), background, backlash, resolution and bandwidth limitations, aberrations, non-linearity, etc. These noise sources are all determined from real measurements and are incorporated in the simulator using semi-empirical functions. The simulator data consists of libraries of synchronized videos or still images of interference patterns, micrometer dial, pressure gauge, digital voltmeter, etc. For each experiment students choose appropriate parameters for scans (start, end, step, dwell), which may require patience and multiple "runs", but builds important experience for "good data-taking". Even though the data are generated by software, the students must await their results for the same amount of time as if the data were being collected by equipment. In some cases, alignment is required. Typically, alignment comprises a 1D or 2D search to maximize a signal. For visual signals (such as interferometer alignment), the user steps through a stack of images to find one correct image amongst 1000 "wrong" ones. A simple "Focus Demo" is available to illustrate the concept.[23] In this example, the two image stacks are: a microscope focus knob animation and a through-focus series of images of the specimen. These image stacks are offset by several frames of hysteresis. Spoiler: the user can dial too far and crash the microscope slide. Fortunately, this is reversible in the simulator, unlike in real life. For the experiment simulators, "alignment errors" do not propagate, thus students can always obtain "good data". This is not always true for in-person labs, much to the chagrin of students (and instructors).

The simulator modules have mostly been coded as Excel Visual Basic (VBA) macros, which run only on Windows machines. In some cases, the VBA code has been ported to HTML5, which runs via browser on "any platform" and is well-suited for server-deployment with large classes.

### *Example: Interferometry*

The interferometry experiment is based on the Pasco "Precision Interferometer" apparatus. It features a HeNe laser source and components that can be configured on a 30x30cm optical table to build a Michelson or a Fabry Perot interferometer. Students mount and align components, then use "fringe counting" to measure the refractive index for air and for glass. We have added a spatial filter and a CCD linescan camera to record ring-pattern intensity profiles.

Students begin by watching a short video showing an expert instructor running the real experiment in the lab, which provides context for the simulator. That video is available at [24]. A short quiz tests comprehension. The video contains narration, close-ups, still images, etc that illustrate real lab features such as: scale, layout and assembly of the apparatus, cautions for working with lasers, exploring the beam location, shape and blanking using an index card, tips for complex alignment (spatial filter, for example), precision tapping alignment of "cheap" mechanical components, etc. Next, they measure critical dimensions (with uncertainties!) of the apparatus from images with an embedded ruler, using ImageJ software.

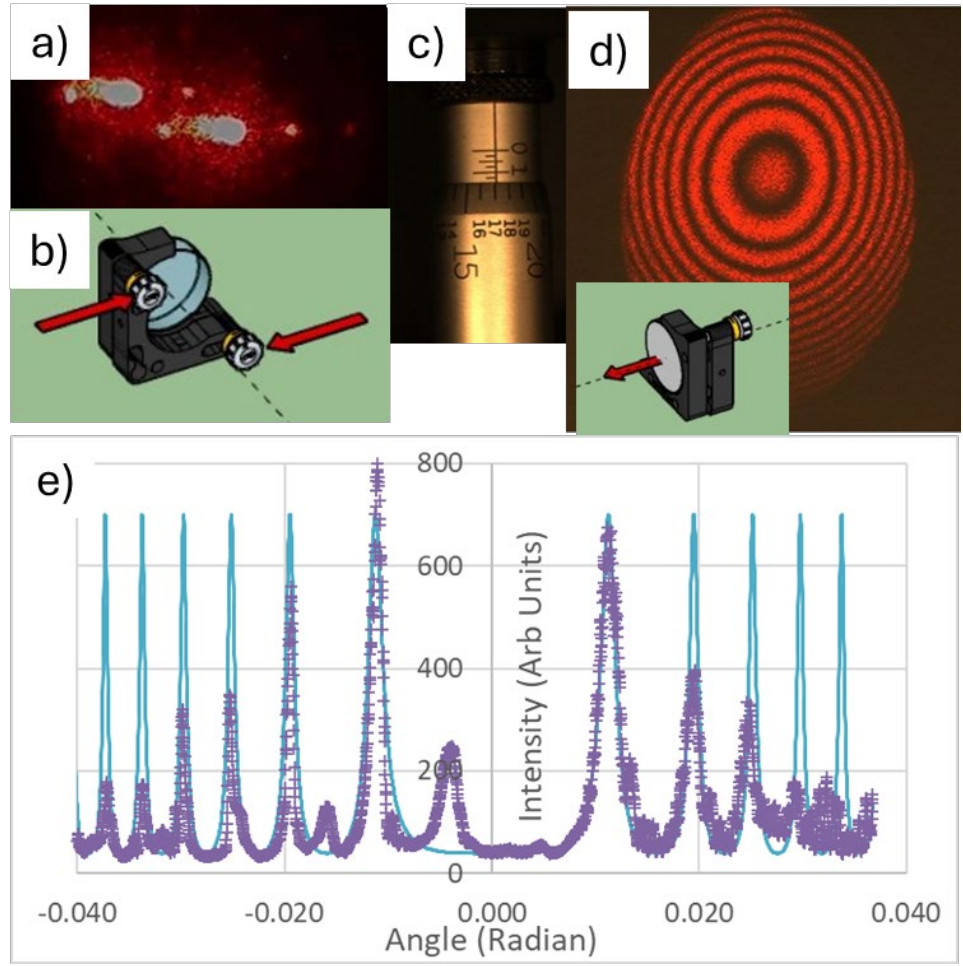

Fig 1. Simulator components for interferometry showing: a) spot-triplet; b) X-Y tilting mirror; c) micrometer; d) ring-pattern and associated translation-mirror (inset); and e) Linescan and fit for Fabry-Perot ring pattern.

Figure 1 shows components and data results for the interferometry simulator. Students perform a 2D alignment with an X-Y tilt mirror to bring two spot triplets into coincidence. They also learn to read a micrometer vernier scale and account for backlash, which is included in the simulator. Next, they measure refractive index for air, counting fringe-shift as air is leaked into an evacuated cylinder. They also compare results for helium and argon gas (bonus, not available in the in-person class).

Next, the optical components are reconfigured to build a Fabry-Perot interferometer with convergent illumination to form concentric-ring patterns. The linescan intensity profiles are fit to

$$I(\theta) = I_0 \ [1 + F\sin^2(\pi m\cos(\theta))]^{-1}, \tag{1}$$

with finesse coefficient $F=4r^2/(1-r^2)^2$, where r=mirror reflectivity amplitude, m=2d/λ, d = mirror separation and θ = angle from the pattern center. The fit yields r= 0.90±.05 and m= 7900±100. This is an interesting example in several aspects since the students must: Do a manual fit due to the cyclic nature of the function; Consider peak positions but not intensity; Ignore stray peaks which arise from ghost reflections from the back side of the mirrors. Students also make a direct measurement of mirror reflectivity by removing the spatial filter (leaving the collimated laser beam) then mis-aligning the mirror tilt to spread the spots. Adjacent spots are attenuated by factor $r^4$, since each mirror in the over-back trip reflects intensity $R=r^2$. Fitting this data yields $r = 0.88\pm.04$. This "cross-examination" value compares nicely with "r" obtained from the ring pattern lineshape.

***Example: X-Ray Physics:***

The apparatus for the X-Ray Physics experiment is a commercial product from Leybold-Didactic (Klinger Scientific in USA) designed for teaching. It features: an x-ray source (35kV, 1mA) with molybdenum anode; a flourescent screen and viewing camera for x-ray projection images and tomography; and a goniometer to control angles for the sample and detector.

Students first view a brief video showing the actual experiment with an expert instructor, to provide context for the virtual lab. The experiment begins with simple "X-Ray Imaging" with a fluoroscope. In the simulator, students select a specimen (Einstein model or Al wedge), turn on the beam and watch the image form gradually during a 5 sec exposure. Examples are shown in Figure 2 below. For the "Einstein Exam", students calculate the radiation "dose equivalent" (in REM, derived from x-ray intensity of 20mW-cm$^{-2}$) and compare that with typical values in the real world.

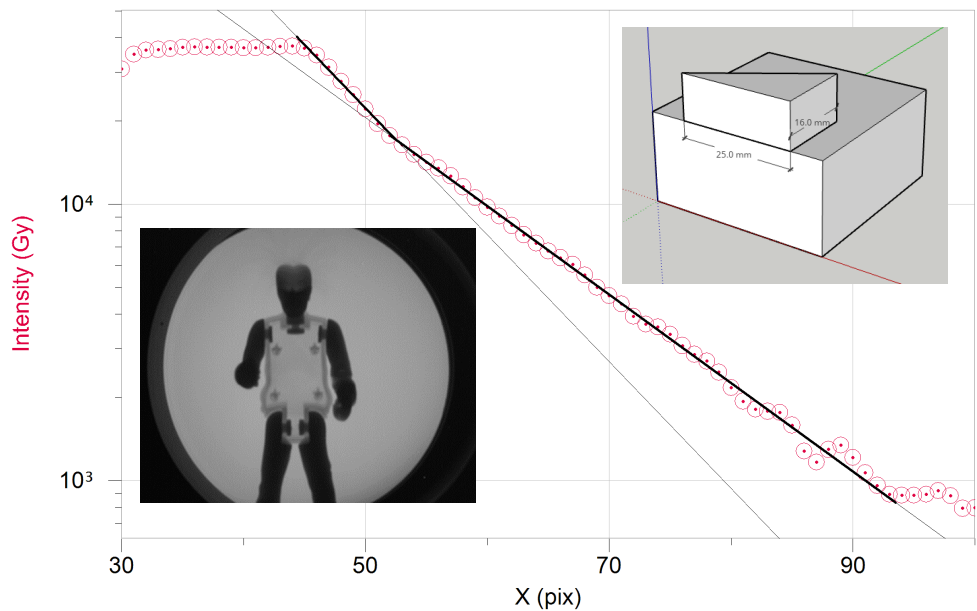

Fig 2. Plot of x-ray intensity vs. horizontal position for Al wedge sample showing two regions with different attenuation. Insets show fluoroscope image of patient Einstein, and perspective drawing of Al wedge sample.

The Al wedge sample is used to make a quantitative measurement of attenuation vs depth. The wedge shape facilitates that in a single image, since the depth is linear with distance from the sharp edge. The image is saved in 16-bit linear format. Students utilize ImageJ to make horizontal Plot Profiles at locations above, centered and below the wedge, which yield incident intensity, attenuated beam and background level (behind a thick pedestal), respectively. The intensity profile data are fit to an attenuation function

$$I(y) = I_0 \exp(-\mu y).$$ (2)

It is apparent that μ varies with the depth "y" (not a straight line on semi-log plot). We show two fit regions with values $\mu_1 = 5.6 \pm 0.2 cm^{-1}$ and $\mu_2 = 3.9 \pm 0.02 cm^{-1}$ for ranges X=45-53pix and 53-93pix, corresponding to 0-1.6 mm and 1.6-8 mm from the sharp edge, respectively. Distance units are converted using 85 pix = 25mm horizontal = 16mm depth. The variation of μ arises because the incident radiation (mostly Brehmstrallung) is broadly spread from 15-30 keV, and μ is strongly dependent on energy ($\mu \propto E^{-3}$). The fit values compare with known values μ ~ 5 cm$^{-1}$ at 25 keV and μ ~ 1 cm$^{-1}$ at 15 keV. Students work out these comparisons using NIST tables for mass-attenuation coefficient μ/ρ ~ 2 cm$^2$/gram for Al at 30keV.

The next section of the experiment examines Bragg scattering from a sample of NaCl. Entrance and exit slits are added (1mm at 5 cm) and the goniometer is set for "coupled θ-2θ scan". In this configuration, the apparatus serves as a monochromator, yielding an energy spectrum for the radiation from the x-ray tube. This follows from Bragg's Law

$$2dsin(\theta) = m\lambda,$$ (3)

where d = 2.82nm is the NaCl lattice spacing, λ is the x-ray wavelength, m (integer) is the order of harmonic and θ is the glancing angle from the surface. The intensity scan comprises an energy spectrum (with higher energy to the left) since $E = hc/\lambda$.

A screen shot of the simulator for Bragg scattering is shown in Figure 3. Cells with yellow highlight allow user- controlled values for the scan, with a slider for beam energy (20-35 keV), and drop-down selectors for beam current and slit size. The scan is constrained to "coupled θ-2θ" mode, which is shown with a live animation for sample and detector angles. The simulator spectrum matches the physical lab spectrum for intensity, position and width of features using semi-empirical formulas. Poisson noise is added with √N standard deviation.

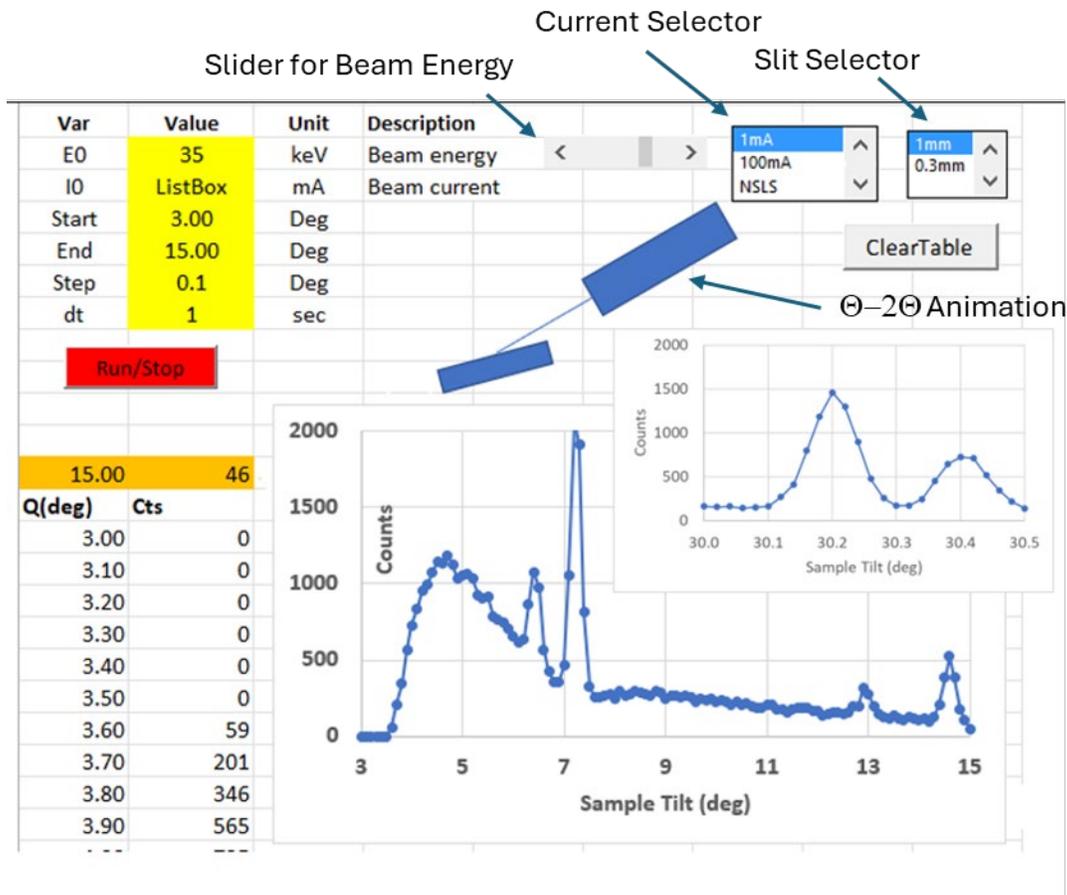

Fig 3. Simulator for x-ray physics experiment showing user-inputs (beam energy, current, slit) and intensity-angle scan at low resolution (main figure) and high-resolution scan (insert) showing m=4 harmonic for $K_\alpha$ line, with doublet structure.

For example, in the setup scan shown in Figure 3, the broad background is Brehmstrallung radiation, with a sharp onset near electron beam energy of 35keV, broad peak near 28keV, and a continuous tail at lower energy (higher angle). The sharp peaks are $K_\alpha$ and $K_\beta$ atomic core levels for the anode material, molybdenum. Students will change e-beam energy and see the Brehmstrallung background shift accordingly, while the atomic line positions remain fixed. The spectrum contains higher-order harmonics with reduced intensity at higher angles, following m=1,2,3 in Bragg's Law (Eq 3). The harmonics are a nuisance but can be utilized for "enhanced resolution" since the splitting or width of peaks increases in proportion to "m".

Students run several scans to locate peaks and to understand the ubiquitous trade-off of count rate vs resolution and noise. For example, the setup scan shown here (0.1 deg step, 1 sec dwell) runs for 2 minutes. A repeat scan with 10 sec dwell runs for 20 minutes, increasing the signal level by 10x and improving the Poisson signal/noise by a factor √10. The high-resolution scan (Fig inset: 0.3mm slit, 0.02 deg step, 100sec dwell) shows the m=4 harmonic for the $K_\alpha$ line, showing clearly resolved peaks at 30.2 and 30.4 deg which are not apparent in the low-resolution scan (Main Fig: single peak at 7.2deg). The narrower slit yields a lower background and narrower peak width but unchanged maximum intensity (assuming perfect alignment). This scan runs for 1 hour. After completing this exercise, students are invited to submit a 1-sentence proposal for beam time at the National Synchrotron Light Source (NSLS), where much higher intensity is available. Technically, one does not have a molybdenum anode source at NSLS, but the behavior of improved signal/noise for high intensity is valid. The proposal review garners a password to unlock high beam current, and the scan can be completed in seconds.

### *Example: Quantum Entanglement:*

The Quantum Entanglement lab is based on a Quantum Entanglement Demo (QuED) kit from QuTools, which we utilize in our in-person courses. This apparatus produces entangled photon pairs via Spontaneous Parametric Down-Conversion (SPDC) process, with multiple cooled coincidence-counting detectors. Signals are routed via fiber-optic cables allowing convenient and safe operation in a teaching lab environment. Our setup includes modules for a Quantum Eraser experiment, a Hong-Ou-Mandel measurement and Quantum Key Deployment via BB84 protocol. Similar kits are available from several vendors at this time (Qubitekk, Thor Labs, Quantum Design North America). A number of groups have built SPDC entanglement experiments "from scratch", following detailed descriptions from Dehlinger et al.[25]

Our simulator closely replicates the QuED experiments above, including alignment, scanning, polarization, coincidence counting, etc with realistic count rates and "visibility". An example is shown in Figure 4. User inputs (yellow highlighted) include: laser intensity (max 40mA), coincidence window (min 10ns), dwell time, optional offset delay for counter "B", and blanking for each detector. The experiment selector includes 3 choices: beam-splitter, SPDC twins and heralded beam-splitter (Hanbury Brown Twiss) and the schematic layout for each case updates when selected. Counts are tabulated for detectors individually and in coincidence, with 10 runs in a set, and the top row (shaded tan) shows the sum of 10 runs. This simulator layout reminds students of the statistical variation in counts during repeated measurements. For the beam-splitter experiment, students measure coincidence counts (BB') vs coincidence window $\tau_c$. This shows that the BB' counts are "accidental", giving the conclusion that a photon passing through a beam-splitter appears at either B or B' with equal probability, but never both. The SPDC experiment, by contrast, shows that the SPDC twins do arrive in coincidence (though at only ~ 5% of the single rate due to detector efficiency). The HBT gives experimental proof that the observed photons are indeed single photons (a non-trivial result). This entire exercise is beautifully described in Pearson et al, which provides an excellent guide to background, theory and interpretation.[26]

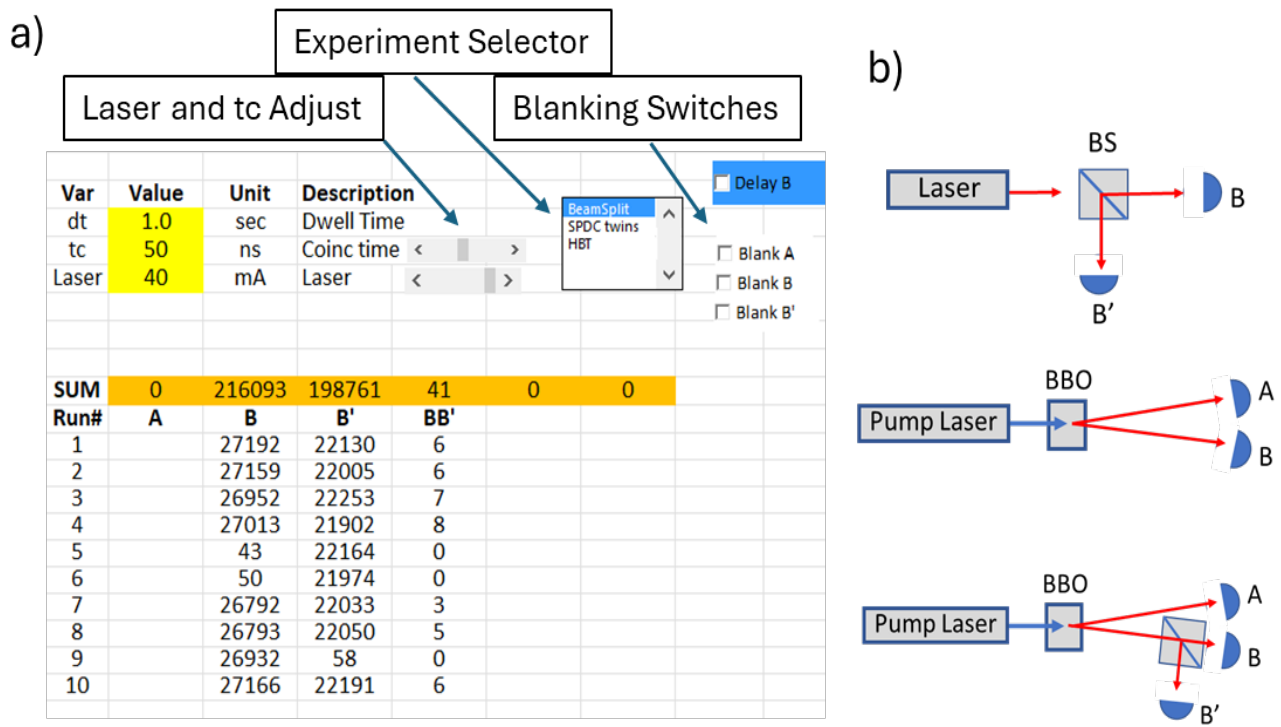

Fig 4 Simulator for Quantum Entanglement experiment showing: a) Counting experiment for beam-splitter with laser adjustment, experiment selector and blanking switches and b) Schematic layouts (update with selector setting) for BeamSplitter, SPDC twins and HBT experiments.

***Cohort and Performance:***

The author has designed, built and delivered the Advanced Lab courses at ASU, first with in-person format (10 years) and later with online format (10 years). Within those groups, the average performance of the online cohort was comparable to that of the in-person cohort. The online cohort, however, had a broader distribution, with some students unable to pass the course in their first attempt and others who demonstrated skills and understanding beyond the best of the in-person cohort. Some online students bring years of experience in technical fields working on large projects with teams of co-workers. It is very gratifying to see them dig into these courses and usefully augment their skills and interests.

The growth in ASU's fully online physics BS degree, since its inception in 2021, has been stunning, with current enrollment (Spring 2025) of 478, compared with 178 for the in-person program. The first cohort of online students is just graduating, and a significant number of students have applied to PhD programs in physics. It is too soon to measure their acceptance rate. Reference letters from the author often begin by explaining the nature of our online degree, especially the Advanced Lab courses. Colleagues in related departments/fields (electrical engineering, bio-chemistry) with fully online degrees at ASU report good success with placing students into graduate programs.

### Summary and Conclusion:

In this paper we have described the structure and content of the online Advanced Lab courses within the fully online BS degree in Physics at ASU. We believe this provides a "proof of concept" for online Advanced Labs in general and will encourage their development and use elsewhere.

### Supplementary Materials:

In the Supplementary Materials section we provide a complete collection of all items comprising our online Advanced Lab courses, including tutorials, handouts and simulators with accessible code. Print readers can see these materials at [DOI link inserted by AIPP]. All materials are licensed as Creative Commons CC-BY-NC, which allows non-commercial public use and adaptation with proper citation.

### Acknowledgements:

We gratefully acknowledge contributions from EdPlus, the ASU online college, particularly Mary Loder, instructional designer, Chandima Cumaranatunga, programmer and Ricardo Leon, videographer.

***Appendix: Sample Assignments***

In this appendix we give 3 sample assignments so the reader may judge their complexity and context without venturing into the Supplemental Materials documents. The links below in this document are not active. Assignments are due essentially every class day. They range in difficulty from 30min to several hours. Note that background reading or web-page or tutorial videos are often included in the assignment. Each assignment is graded within 24 hours with a numerical score (small integer) for each item (5, 3, 3 items in the examples below), plus occasional comments if appropriate. In the PHY334 course there are approximately 150 "scores" in total. We provide a suggested scoring for all items in each module.

***Length measurement using ImageJ:***

You will occasionally need to make measurements of distance/length from images of the apparatus. You will use ImageJ, which is a powerful, open-source (free) program that runs on multiple platforms (Windows, iOS, etc). It is widely used for high-end research and is very powerful, particularly with the addition of user-defined modules. The on-line help is very good and should be consulted for details.

    a. Download ImageJ and install it.
    b. View our tutorial at ImageJ Profile Tool.
    c. Run ImageJ and open the image MeasureLength.jpg.
    d. Measure the pixel length for a line drawn on the ruler to find the calibration (pixels/mm) from this image, along with uncertainty. You should choose a wide span near the image-center. Include the image with your linescan in your "answer".
    e. Find the width of the tape-rule (mm), with uncertainty.

**Micrometer:**
Note that the circular fringe pattern collapses/expands as you turn the micrometer (**mic**). The mic translates mirror M1 along the beam via a lever mechanism that reduces the physical motion by some factor, which you will determine in this exercise. The motion of the RingPattern (**RP**) is cyclic, with a period of 1 fringe when the round-trip path length changes by distance $\lambda$. That is, one cycle of the RP corresponds to a M1 displacement of $\lambda/2$.

a. Find the micrometer reading (0.xxx with units) for the image MicScale.jpg. This is an Imperial mic (inches), as is most common in the USA. Instructions for reading a mic are available at How to Read a Micrometer.
b. Move the frame counter FWD and BWD to find the "backlash" (# frames).
c. Find the lever reduction factor $f$ = (mic displacement)/(M1 displacement) by counting fringe-cycles for a certain mic displacement. For example, record mic values corresponding to 10 cycles of RP.

**Refractive index for gases:**

This assignment requires running the simulator to obtain data. The questions below follow a detailed explanation of concept (suppressed here), with apparatus diagrams and equations. Measure the fringe shift (count) vs pressure, as you bleed gas into the cylinder.

a. Plot this data and make a linear fit to n(P), as Eq3. From this, find n(Pa), the index of refraction at atmo pressure. Note: Since n for gas is only slightly larger than 1, that is $n = 1+\varepsilon$, with $\varepsilon \ll 1$, it is useful to work with (n-1) rather than n.

b. Repeat the procedure for He and Xe. Note that "$\varepsilon$" tends to scale with "polarizability", which also scales with physical size of the molecule.